# Research Strategy and Scoping Survey on Spreadsheet Practices

Thomas A. Grossman
OPMA, Haskayne School of Business, Calgary, Alberta, Canada T2N 1N4
Thomas.Grossman@Haskayne.UCalgary.Ca
Ozgur Ozluk
ISBA Department College of Business, SM San Francisco, CA 94132
ozgur@sfsu.edu

**ABSTRACT**

We propose a research strategy for creating and deploying prescriptive recommendations for spreadsheet practice. Empirical data on usage can be used to create a taxonomy of spreadsheet classes. Within each class, existing practices and ideal practices can he combined into proposed best practices for deployment. As a first step we propose a scoping survey to gather non-anecdotal data on spreadsheet usage. The scoping survey will interview people who develop spreadsheets. We will investigate the determinants of spreadsheet importance, identify current industry practices, and document existing standards for creation and use of spreadsheets. The survey will provide insight into user attributes, spreadsheet importance, and current practices. Results will be valuable in themselves, and will guide future empirical research.

## 1. INTRODUCTION

An important question facing the spreadsheet research community is what must be done to influence the way that industry creates, uses, and manages spreadsheets. In particular: How can we develop and deploy prescriptive recommendations for effective spreadsheet practice?

A major barrier to answer this question is the enormous diversity with which spreadsheets are used, which makes it difficult to generalize about spreadsheets. To solve this problem [Grossman 2002] called for research on developing a "taxonomy" to allow classification of spreadsheets. The creation of this taxonomy is no easy task. We present a research strategy for developing a taxonomy and using it as the foundation for prescriptive research. As a first step toward developing the taxonomy, we propose a focused survey of spreadsheet usage.

## 2. RESEARCH STRATEGY

There is wide variation in spreadsheet usage, and corresponding wide variation in potential opportunities to make usage more effective. Because of this variety, it is difficult to develop useful generalizations or theories regarding spreadsheets. Procedures that are effective for one set of spreadsheet users cannot reliably be applied to others.

We propose a research strategy (Figure 1) to guide future research. This strategy uses empirical research on spreadsheets, to formalize our understanding and enable us to create a taxonomy of spreadsheets. This taxonomy will organize and simplify the messy world of spreadsheets, and provide the foundation for prescriptive practice recommendations.

The base level of the strategy is the universe of spreadsheets, represented by the box "Spreadsheets in the World". The next box up represents our "Understanding of Usage", which currently is largely anecdotal and not well organized. The next box up is a set of organizing principles, called a "Taxonomy". The Taxonomy consists of many "classes", where each class is a set of spreadsheets with similar characteristics. We can think of the Taxonomy as a map of "Spreadsheets in the World".

Each class in the taxonomy has an associated set of practice issues, indicated by the dotted-line boxes. For each class, we will understand the empirical issues of "Existing Practices", and devise a set of principles called 1deal Practices". We will synthesize the two into "Proposed Best Practices". For the Proposed Best Practices to be meaningful, they must be transferred to industry, and therefore we need to consider their "Deployment". The double-headed arrows indicate interaction among boxes.

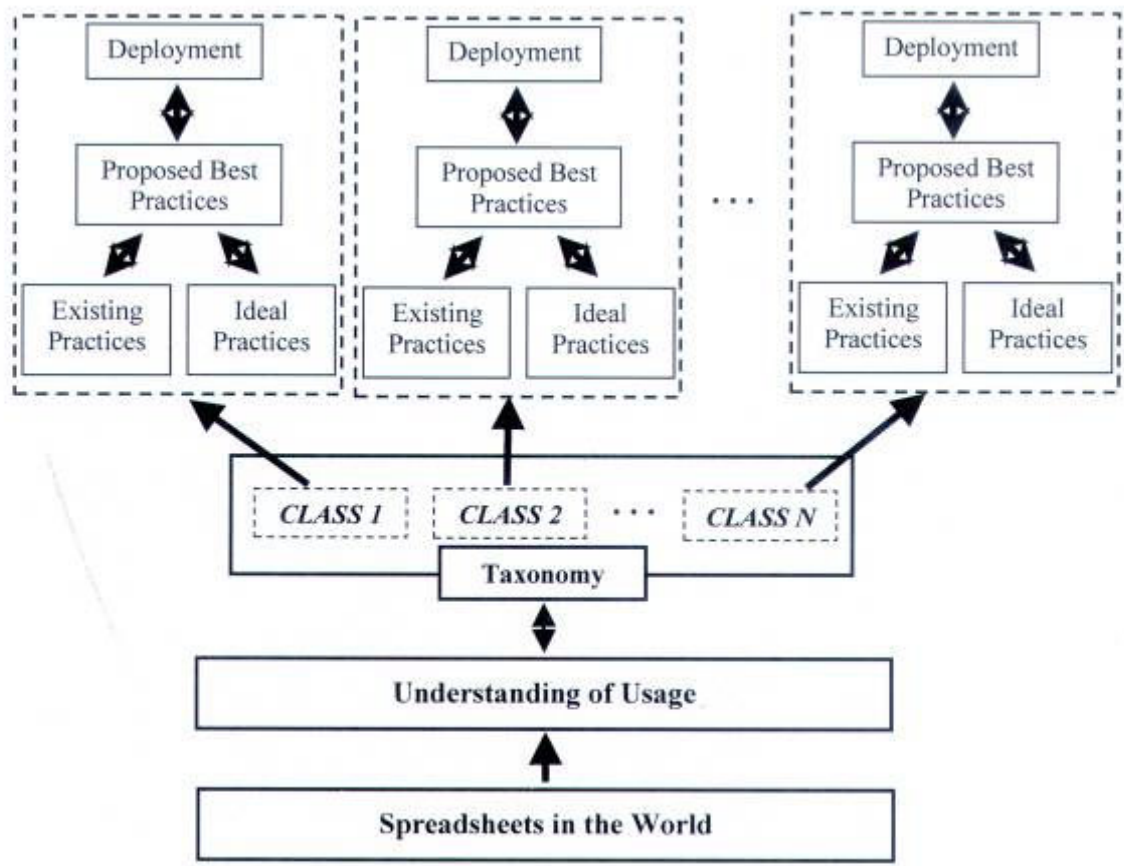

Figure 1. Research Strategy

## 2.1. Spreadsheets in the World

Spreadsheets have enormous diversity. They are used by millions of people in virtually every industry in every country. They serve many different roles, range in size from tiny to 100's of megabytes and millions of cells. They are created, used, distributed, and modified in myriad different ways.

This diversity is evidence of spreadsheets' profound, even revolutionary success. But this diversity makes it very difficult to say anything useful about spreadsheets in general. We must carve out pieces of the spreadsheet world that are small enough to examine in detail, starting by understanding usage.

## 2.2. Understanding of Usage

Our current understanding of usage is mostly anecdotal. We need rigorous empirical research to generate reliable data on spreadsheet usage. This research will explore and document the variety of ways in which spreadsheets are created and used. To fully understand spreadsheet usage will be the work of many years.

There are many empirical questions to be considered. For example, Who are spreadsheet users? How are spreadsheets developed? How are spreadsheets used? How are spreadsheets managed? What is the importance of spreadsheets? What is the impact of errors?

To get started, we propose in section 4 a survey of usage. This survey, and its successors, will provide data on important basic facts about spreadsheets, and help us to identify possible taxonomic classes.

### 2.3. Taxonomy

A **taxonomy** organizes spreadsheets into different "classes". Within a class, spreadsheets share similar characteristics. Two examples of classes might be "personal productivity tools developed by end-user modelers for one-time use" or 'Spreadsheets used to support allocating multi-million dollar advertising budgets across products in a large company". ([Huff and Rivard 1985] survey end user developers to devise taxonomies of such users and the applications they develop.)

The taxonomy will serve as a "map" to the "spreadsheet world", and provide a foundation for spreadsheet research. The power of the taxonomy is that within a class it should be possible to develop and test theories about spreadsheet that can be generalized to all spreadsheets in that class. Therefore, developing a useful taxonomy is a high priority.

The choice of classes is an important open question. Classes could be defined based on *intended audience* (written for self or for someone else), *programming attitude* (amateur or professional programmer), *activity* (what a spreadsheet user does), or other attributes. Classes could also be defined by industry, function, experience of user, type of application, problem being solved, or other dimensions.

We can move spreadsheet research forward with only a few classes, provided they allow us to set boundaries for statements about practice. It is not necessary to define a "best" taxonomy, or even a complete taxonomy.

### 2.4. Spreadsheet Practices

Within a class, we are interested in spreadsheet practices. **Practices** include all aspects of how spreadsheets are created and used. There are many issues that could be considered. These include: lifecycle models; how a quantitative model implemented in the spreadsheet is developed; processes, procedures, or standards that are employed, whether formally or informally; how spreadsheets models are tested, verified and validated; how spreadsheet quality is evaluated; how datasets are passed through spreadsheet models; how spreadsheets or spreadsheet outputs are transferred to other people; and how changes are made to existing spreadsheets.

As shown in Figure 1, we work within a class by obtaining an empirical understanding of "Existing Practices". Drawing from many sources, we will develop theoretical recommendations for "Ideal Practices". By adapting ideal practices to the exigencies of the real world, we will develop a set of recommended "Proposed Best Practices". Proposed best practices might be testable in the laboratory or in the field. Ultimately, we would need to develop techniques for **Deployment,** transferring techniques to industry so that they are adopted and used by spreadsheet developers in that class.

### 3. EXPLORING SPREADSHEET USAGE: TOWARDS AN INDUSTRY SURVEY

There has been much discussion in the spreadsheet research community regarding a survey of spreadsheets in business. To our knowledge, no broad survey has been published. [Klein and

- 3 -

Myers 1999] provide principles for conducting and evaluating interpretive field studies in information systems.

[Cragg and King 1993] performed an exploratory survey of spreadsheet practices in ten firms, with an emphasis on the process of building spreadsheet models. The study showed that spreadsheet models were usually built in an informal, iterative manner, by people from all organizational levels, with very little training in the building of models. At least 25% of the models contained errors. They argue that observed spreadsheet practices are inadequate and call for increased training as well as setting and enforcing organizational spreadsheet standards.

[Hendry et al 1994] interviewed users regarding key strengths and weaknesses of spreadsheet models. They found that strengths related to gratification of immediate needs, while the weaknesses make subsequent debugging and interpretation difficult, "suggesting a situated view of spreadsheet usage in which present needs outweigh future needs". [Nardi and Miller 1990 1991] performed an ethnographic study of how users worked collaboratively to solve problems using spreadsheets. [Seal et al 2000] provide a survey of spreadsheet management science applications.

A discussion document has been circulated by Mike Racer ([mRacer@mocha.memphis.edu](mailto:mRacer@mocha.memphis.edu)), with input from Ray Butler, Steve Powell, Tom Grossman and perhaps others. This document has helped us understand that we are profoundly ignorant of spreadsheet usage, and that there are an overwhelmingly large number of aspects of spreadsheet usage that are unexplored.

We believe that an important first step is to undertake a small, scoping survey. We intend the results to be useful in themselves, and informative for future empirical research. Through this scoping survey, we hope to obtain valuable data on current practices across a broad range of users. This survey will interview a small number of subjects for two reasons. First, a small number of interviews will facilitate an in-depth analysis. Second, we are beginners at field research, so we should start small.

## 4. PROPOSED SCOPING SURVEY

The scoping survey is intended to provide data on the usage, importance, and development practices for spreadsheets. We present our current thinking regarding what makes most sense. We hope that conference participants will help us improve this survey.

### 4.1. Spreadsheets Considered

In this survey, we ask individuals to select a few spreadsheets (perhaps 3 or 5) that they deem to be "important". We will ask questions about each spreadsheet to gather data. Analysis of the data should lead to useful generalizations and insights into spreadsheet usage.

We will study only "important" spreadsheets. The definition of "important" will be made by the research subjects. Our data will include the subjects' explanation of why those spreadsheets are important. We limit this analysis to "important" spreadsheets for several reasons.

There is no documentation in the literature that spreadsheets are important to industry and therefore merit serious attention. Therefore, we need data on "importance". Second, we believe that "important" spreadsheets are more likely to have procedures or standards than unimportant spreadsheets.

### 4.2. Research Goals
This research has the following goals:

- Investigate the determinants of spreadsheet importance.
- Identify current practices in industry for creation, use, dissemination, and management of spreadsheets
- Document existing standards for creation and use of spreadsheets

### 4.3. Investigative Questions

At the end of our survey, we will have data on the following topics:

A. <u>User Attributes:</u> Who creates spreadsheets?
   Survey Data: Description of creators experience, education, training, and so on.
B. <u>Spreadsheet Importance:</u> What causes a spreadsheet to be deemed important by users?
   Survey Data: *For each spreadsheet:* Determinants of importance. Ways to describe or measure importance. Features of important spreadsheets.
C. <u>Current Practices:</u> What practices are employed to create and use an important spreadsheet?
   Survey Data: *For each spreadsheet:* Formal or informal standards or habits used and their origins. Any common software engineering principles used. The description of spreadsheet design process. The impact of dissemination on the spreadsheet. Accuracy evaluation techniques.

## 5. SPREADSHEET USER CLASSIFICATION:

We need to classify different roles people take when they use spreadsheets. In classic software engineering, an "analyst" works with "users" to develop specifications. The specifications are provided to "developers" who create the software. With spreadsheets, these traditional demarcations do not seem to apply. Instead, there is a strong tradition of end-user development, where the person with a problem develops a spreadsheet for their own use. Sometimes, this spreadsheet is provided to other people, and end-user development becomes development for others. Other people may use the spreadsheet without changing it or may adapt it to be their own. In many cases, the outputs from the spreadsheet (charts, numbers, etc.) are provided to people who never see the spreadsheet.

We propose the following classification of spreadsheet users:

<u>Spreadsheet Developer (SD):</u> Designs, develops, programs, tests spreadsheets. Includes end-user developers, traditional developers, and people who further develop another developer's spreadsheet.

<u>Spreadsheet Receiver (SR):</u> enters data, examines results, can give feedback about structural aspects of the spreadsheet to the SD, but does not program the spreadsheet.

<u>Spreadsheet Consumer</u> (SC): observes results from spreadsheet models, but not the spreadsheet itself. Can give feedback about desired spreadsheet outputs to the SD and SR

*For this scoping survey, we will concentrate solely on Spreadsheet Developers.*

## 6. SURVEY QUESTIONS

We will categorize our survey questions into three groups, depending on which investigative question they are answering. Survey questions are either questions for which we look for specific answers from the research subjects or questions in which we would like to get a free form answer in their own words (labelled with "FFA" below).

### 6.1. Questions on User Attributes

The following questions are necessary to learn more about the spreadsheet developer's background on spreadsheets.

- What training have you received in creating spreadsheets?
- How did you learn to create and use spreadsheets?
- How do you improve your spreadsheet skills?
- What % of your time do you spend working in spreadsheets?
- How long have you been preparing spreadsheets?
- How would you rate your skill at spreadsheets?
    - SCALE 1-5, from Novice to Expert.
- Part of my professional identity is my skill with spreadsheets.
    - SCALE 1-5, Strong Disagree-Strong Agree.
- What computer languages and operating systems have you used?
- What training or experience do you have in software engineering?
- Have you over opened one of your own spreadsheets, and had difficulty understanding it?

## 6.2. Questions on Spreadsheet Importance

One of our goals is to determine what makes a spreadsheet "important" in the business world. To reach this goal, we first determine how the importance of a spreadsheet is evaluated. We then determine the features of this "important' spreadsheet.

We recognize that there are several ways to evaluate importance. For example, importance might be quantified by dollars implicit in the spreadsheet, or monetary benefit from use of the spreadsheet.

Below, we describe different attributes of a spreadsheet and questions to evaluate these attributes.

**Evaluation of Importance**

You have identified this spreadsheet as important. Why? (FFA)

**Management**

What would the impact be if this spreadsheet were irretrievably lost? (FFA)What would happen to this spreadsheet and the information it produces if you suddenly became unavailable? Do others know how to access it? Do others know how to use it? (FFA)

**Size**

What is the size of this spreadsheet? Quantified by:

- Number of non-empty cells
- Number of cell formulas
- Size of file

We may want to divide these into categories, such as Small, Medium, and Large. Initial thinking makes us believe that the larger a spreadsheet gets, the more attention needs to be given to the creation process.

**Intended Audience**

- Who is the intended audience?

Self            Workgroup            Higher Management            Outside Clients/Analysts
- How well do you understand their requirements? (FFA)
- What feedback do you get from them? (FFA)

The intended audience may affect how a spreadsheet is created, managed, disseminated. Certain attributes, such as accuracy or documentation, may change with the audience of the spreadsheet.

**Frequency of Use**

It is clear that one-time use spreadsheets are sometimes used multiple times. We will investigate this by gathered data on usage frequency. Intended (at time of creation) frequency is "once", "occasional", and "frequent". If possible, we will acquire quantitative data on these. Actual (in use) frequency is an estimate of the number of actual uses in a period of time. The results can be arranged in the table in Figure 2.

Figure 2,     Usage Table

|  |  | Intended Number of Uses | | |
|---|---|---|---|---|
|  |  | Once | Occasional | Frequent |
| Actual Number of Uses | Number |  |  |  |

**Purpose (Motivation for creation of the spreadsheet)**

What motivated you to create this spreadsheet? (FFA)

Anticipated responses include:
- Develop insight or better understanding
- Compute a number
- Answer a question Make a decision Make sense of data
- Communicate about data
- Automate routine task

**Systems Interaction**

A spreadsheet may be part of a larger information system. Its outputs may feed other software, or they may not. A spreadsheet may take inputs from a "live" system, where the inputs periodically change and affect the spreadsheet, or the inputs may be static. We give names to the four combinations as shown below.

|  | No Live Inputs | Live Inputs |
|---|---|---|
| Does Not Feed Other Software | "Stand-Alone" | "Dynamic Analysis" |
| Feeds Other Software | "System Input" | "Integrated" |

**6.3.    Question on Current Practices**

We also would like to probe the Spreadsheet Developers about how important spreadsheets, are conceived, designed, developed, distributed, and managed. The following questions address these issues:

**Development**

How did you decide what to create? (FFA)
Was there requirements analysis?
Was data available before the creation?
Did you have any concerns with the data at any point? (FFA)
How did you decide when to stop? If you had more time, would you have done more work on this? (FFA)

**Design**

How did you start creating this spreadsheet? (FFA) Anticipated answers include the following.
- Design on paper
- Design in spreadsheet
- Conscious prototyping
- Used a well-defined lifecycle model

Have you followed any standards or procedures designing this spreadsheet? If you have, were these:
- Personal/workgroup/corporate rules/standards?
- Formal/informal?
- Describe them (FFA).

**Programming Practices**

Did you use any of the following:
- Labels, Cell Names
- Formats
- Modules
- Multiple worksheets
- Linked spreadsheets

**Distribution**

Do other people have a copy of this spreadsheet? If NO, SKIP. If YES, then ask the following:

- What changes, if any, did you make before giving the spreadsheet to someone else? Why? (FFA)
- Is the spreadsheet maintained centrally, or does each person maintain a copy of their own?

**Accuracy**

Do you believe this spreadsheet is accurate (without errors)?
How confident are you in your evaluation of accuracy?
How did you determine the accuracy of your spreadsheet? (FFA) Anticipated answers might include:
- Outputs made sense
- Systematically audited formulas

- 8 -

- Others evaluated accuracy
- Tests

## 7. INTERVIEW PROCESS

Face-to-face interview with software developer in front of their computer

a. Ask the SD to choose 3 spreadsheets he has worked on in the last six months that he deems to be "important".

b. For each spreadsheet, discuss the importance of the spreadsheet. Ask any questions on Spreadsheet Importance (section 6.2) that were not addressed in the discussion.

c. For each spreadsheet, discuss how the spreadsheet was created and used. Ask any questions on current practices (section 6.3) that were not addressed in the discussion.

d. For each spreadsheet, the researchers visually inspect the spreadsheet, make observations, and compare observations to subject responses to questions. Possible list of aspects the surveyors investigate:
   - Any particular trends observed by the user in the design, layout of the process
   - Surveyor's estimate of the importance of the spreadsheet

e. Ask questions on User Attributes (section 6.1).

f. Ask the SD if there is anything else he/she would like to add to the discussion.